# Simple model of interbank trade with tiering remuneration

Toshifumi Nakamura*

European Central Bank†

May 2019

**Abstract:** Many countries have adopted negative interest rate policies with tiering remuneration, which allows for exemption from negative rates. This practice has led to higher interbank trading volumes, with market rates ranging between zero and the negative remuneration rates. This study proposes a basic model of an interbank market with tiering remuneration that can be tested with actual market data because of its simplicity and can indicate the level of the market rate created by the different exemption levels. By generalizing the model, we found that a tiering system is also suitable for maintaining a higher trading activity, regardless of the level of the remuneration rate.

---

* The views expressed in this paper are those of the author and should not be interpreted as those of the European Central Bank or other institutions the author is working for.

† DG-Market operation, Money market and liquidity division; email: *toshifumi.nakamura@ecb.int*. The author is a secondee from the Bank of Japan; email: *toshifumi.nakamura@boj.or.jp*.

# 1. Introduction

The several central banks that have imposed negative interest rates[1] can largely be classified into two groups. One group uses negative interest rate exemptions to some extent in addition to reserve requirements; this system is called tiering remuneration system. The other group adopts a simpler policy, under which a negative rate is uniformly imposed on all reserves, except for reserve requirements.

For the central banks of the countries that follow a tiering system, such as Denmark, Japan, and Switzerland,[2] a common positive side effect is the increase in the overnight interbank market trading activity (Figure 1), which is a direct result of the arbitrage opportunities arising from the differences in remuneration rates. For example, bank A, which has exceeded the allocated exemption amount wants to lend cash in the market if the market rate is higher than the negative rates charged by the central bank, while bank B, which has not reached the upper limit of the exemption amount, will borrow cash from the market provided that the market rate is below the zero rate.

Figure 1: Market volume of overnight indices

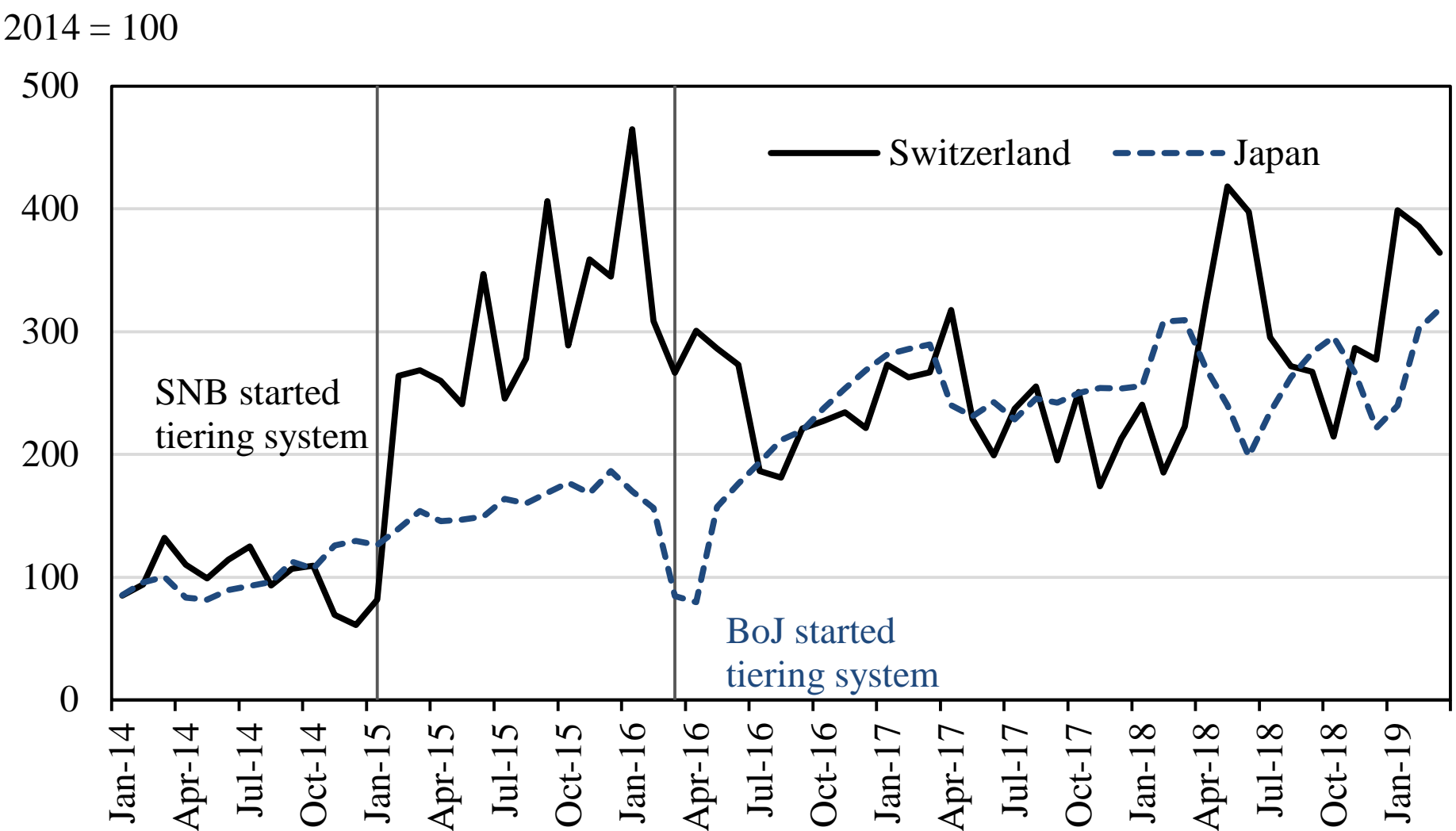

Sources: Six Swiss Exchange, Swiss National Bank (SNB), Bank of Japan (BoJ)

[1] For an introduction on negative rate regimes, see Bernhardsen and Lund (2015) and Bech and Malkhozov (2016).

[2] These three countries are examples of the tiering system by the International Monetary Fund (2017). Jobst and Lin (2016) also refer to Bosnia and Herzegovina, Bulgaria, and Norway. Due to the limitations of market data or the limited effects on the market, we did not include these additional jurisdictions in our analysis.

Therefore, market rates float between a negative remuneration rate and the zero rate, which is the remuneration rate for the exempted amount (Figure 2). Interestingly, the market rate corresponding to the negative remuneration rate differs among the three analyzed countries. Intuitively, this is related to the exemption amount allocated to banks, but a microfoundation is yet to be developed.

Figure 2: Market rate and remuneration rates

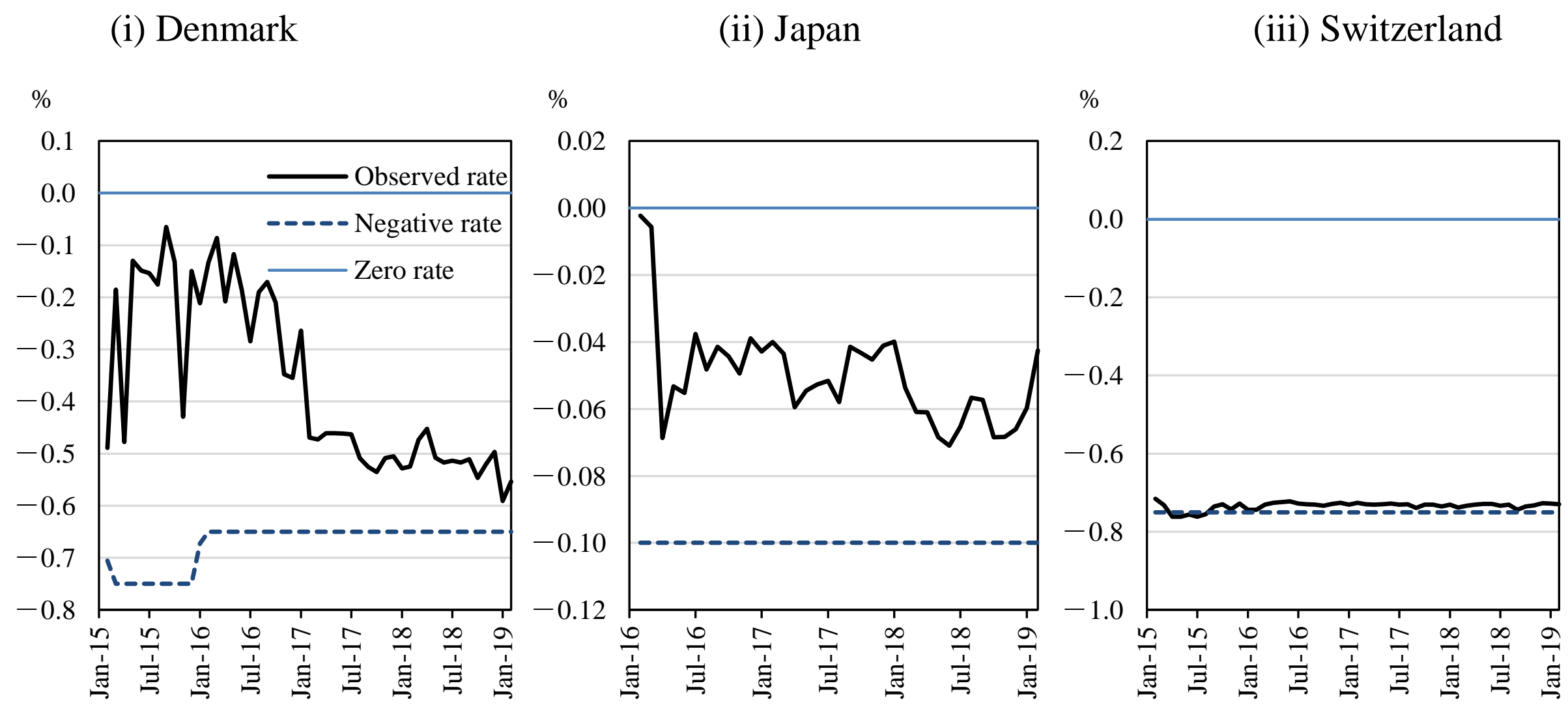

Sources: NASDAQ, Six Swiss Exchange, Danmarks Nationalbank, BoJ, SNB

This study thus formulates a basic model of an interbank market with tiering remuneration and determines a suitable method for rate formation and trading volume. Due to its simplicity, the model can be tested with actual market data. Moreover, it enables us to discuss the probable methods of retaining a good trading activity.

Because this study proposes a simple model for an in-depth understanding of the tiering system, it assumes high liquidity across all banks and negligible payment shocks. Hence, using central bank lending is excluded, unlike in Poole's (1968) framework. If the model were to include such shocks, the market rate would become the risk-weighted average of the central bank's lending and deposit facility rate (Bindseil, 2004). The market rate in our model will be connected to the exemption size.

Recently, Boutros and Witmer (2017) introduced a theoretical model with a tiering remuneration system without testing actual market data. The highly simplified model in this study can be tested by using actual market data. Afonso, Armenter, and Lester (2018) also successfully tested the model by using market data and comparing their model of the United States Fed Funds market with individual bank-level data. However, microdata on other countries are scarce because few countries use the tiering system. Therefore, we used a simplified model, which can be easily tested by the aggregated balances of (a) reserves exceeding exemption and (b) unused exemption amount or the (c) upper limit of exemption and (d) excess liquidity in the system.

## 2. Model and Equilibrium

In this section, a simple static model is developed and the equilibrium calculated.

**Set up for morning.** Bank $i \in [0,1]$ has the profile of excess liquidity over reserve requirements $x_i$ and cost regarding trade transaction (except for the trading rate) $c_i$ (%). We assume uniform and mutual independent distributions of $x_i \sim (0,1)$ and $c_i \sim (0,1)$ for simplicity.

**Tiering remuneration afternoon.** Bank $i$ knows the remuneration schedule before trading: $0\%$ is applied to the reserve position in the afternoon, $x_{\mathrm{a}i}$ up to $u \in [0,1]$, and $-1\%$ is applied to the exceeding amount. The central bank pays $p_i$ to bank $i$ as follows:

$$p_i = \begin{cases} -1 \cdot (x_{\mathrm{a}i} - u), x_{\mathrm{a}i} > u \\ 0, u \geq x_{\mathrm{a}i} \end{cases}. \quad (1)$$

**Trade decision during the day.** Bank $i$ decides to lend $l_i$, borrow $b_i$, or do nothing. The decision is made by comparing market rate $-r$ (%), associated cost $c_i$ (%), remuneration rate $0\%$ for the exemption amount, and $-1\%$ for the exceeding amount.

$$x_{\mathrm{a}i} = \begin{cases} x_i - l_i, x_i > u \text{ and } -r - c_i - (-1) > 0 \\ x_i + b_i, u > x_i \text{ and } 0 - (-r) - c_i > 0 \\ x_i, \text{otherwise} \end{cases}. \quad (2)$$

The concept in (2) is that bank $i$ ($x_i > u$) will lend money if the income from lending $-r$ minus cost $c_i$ minus the opportunity cost to receive remuneration $-1$ is positive. Similarly, bank $i$ ($u > x_i$) will borrow money if the income from borrowing $r$ minus cost $c_i$ is positive.

**Equilibrium market rate.** Market rate $-r^*$ should be decided as to balance the aggregated lending and borrowing needs. This amount is described as trading volume $V$. Further, there are banks that have reserves exceeding $u$ but avoid lending because of cost. The total size of this remaining excess reserve is termed $A$ and the unused balance below $u$ as $B$. Specifically,

$$V = \int_i l_i \,\mathrm{d}i = \int_i b_i \,\mathrm{d}i, \tag{3}$$

$$V = \int_\mathrm{i} \max(\mathrm{x_i} - \mathrm{u}, 0) \cdot 1_{\{-\mathrm{r}^* - \mathrm{c_i} + 1 > 0\}} \,\mathrm{di} = \int_\mathrm{i} \max(\mathrm{u} - \mathrm{x_i}, 0) \cdot 1_{\{\mathrm{r}^* - \mathrm{c_i} > 0\}} \,\mathrm{di}, \tag{4}$$

$$V = \int_i \max(x_i - u, 0) \,\mathrm{d}i\,(1 - r^*) = \int_i \max(u - x_i, 0) \,\mathrm{d}i\, r^*. \tag{5}$$

Since we assume $c_i \sim (0,1)$,

$$A = \int_i \max(x_i - u, 0) \cdot \mathbf{1}_{\{-r^* - c_i + 1 \le 0\}} \,\mathrm{d}i = \int_i \max(x_i - u, 0) \,\mathrm{d}i\, r^*, \tag{6}$$

$$B = \int_i \max(u - x_i, 0) \cdot \mathbf{1}_{\{r^* - c_i \le 0\}} \,\mathrm{d}i = \int_i \max(u - x_i, 0) \,\mathrm{d}i\,(1 - r^*), \tag{7}$$

$$-r^* = \frac{-1}{1 + \sqrt{\frac{B}{A}}}. \tag{8}$$

As we assume $x_i \sim (0,1)$, (8) can be rewritten as

$$-r^* = \frac{-(1-u)^2}{u^2 + (1-u)^2}. \tag{9}$$

## 3. Test with Real Data

This section tests (8) and (9) using market data. The model described in (8) is more general because it is independent of the distribution of $x_i$. However, we need detailed data on tiering, such as the total size of remaining excess reserve, $A$, and the unused exemption balance, $B$. Danmarks Nationalbank releases daily volumes of certificates of deposit, to which $-0.65\%$ is applied, and this is $A$.[3] It also provides current account deposits, to which $0\%$ is applied, as well as the sum of individual current account limits. The difference between these two datasets is $B$. The Bank of Japan also discloses the maintenance period averages for the amount to which $-0\%$ is applied, as well as the amount and upper bound to which $0\%$ is applied. However, the Swiss National Bank does not provide the amount of the unutilized exemption. Alternatively, the trading volume of the Swiss average rate overnight (SARON) is used as input for $V$.[4] Hence, (8) can be tested after rewriting it using $V$ as follows:

$$-r^* = \frac{-1}{1+\sqrt{\frac{B}{A}}} = \frac{-1}{1+\frac{V}{A}}. \tag{10}$$

For the observed market rates to be compared with $-r^*$, one-day rate indices are used, such as Denmark's tomorrow/next rate, Japanese unsecured overnight call rate (TONAR), and SARON.

The necessary data to test (9) are the upper limit of exemption $u$. Assuming a bank's reserve position in the morning is uniformly distributed $x_i \sim (0,1)$, $u$ can be described using the exemption share compared to the aggregated liquidity in the system $ES$ as follows. Further, $ES$ is available in all three countries.

$$ES = \frac{u}{\int_i \; x_i \, \mathrm{d}i}, \tag{10}$$

$$u = \frac{ES}{2}. \tag{11}$$

---

[3] It attributes lower trading volumes to the size of liquidity (Danmarks Nationalbank, 2018), which can be explained by the sizable $A$ and small $B$.

[4] Using market trading volume data $V$ is not always desirable because of its limited coverage compared to the total size of interbank lending. However, the Swiss National Bank mentioned in an annual report that "exemption thresholds have been almost entirely exhausted for some time now" (2018); this is consistent with the estimated figure of $B/A \approx 0.0001$ using $V$.

Figure 3 shows a plot of the quarterly average data on $B/A$ & $ES$ and the observed market rate for the three countries. The observed rate is standardized by dividing it by the degree of negative remuneration (0.65%, 0.10%, and 0.75% for Denmark, Japan, and Switzerland, respectively). The line shows the theoretical $-r^*$ using (8) and (9) + (11). Despite the oversimplification, the real data do not differ much from theory, at the least in approaching the theoretical value.

Figure 3: Relationship between observed market rate, $B/A$, and $ES$

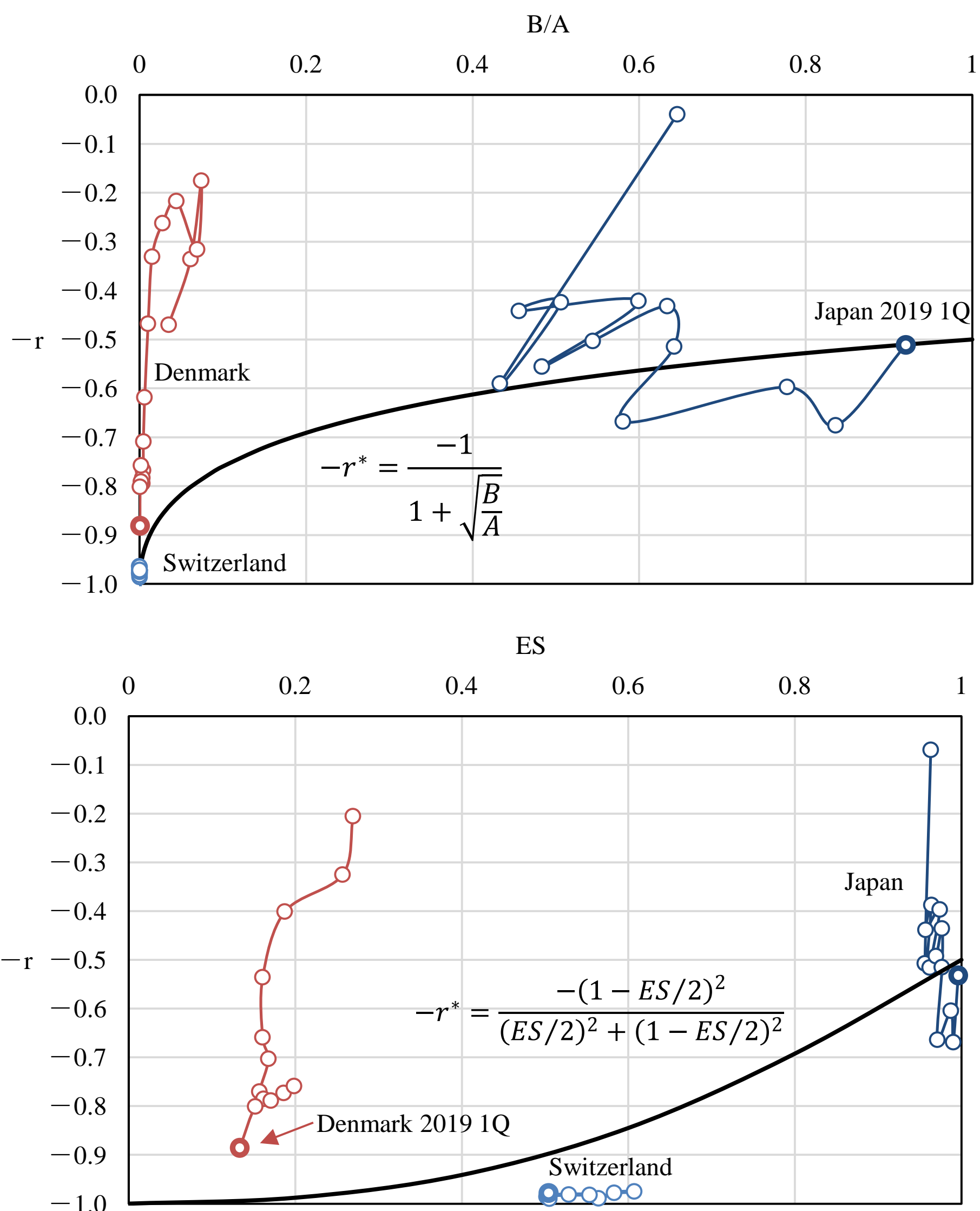

Sources: NASDAQ, Six Swiss Exchange, Danmarks Nationalbank, BoJ, SNB

Moreover, when we focus only on the direction of the move of the theoretical and real market rates, both are moving in a similar manner, as shown in Figure 4. Therefore, the real mechanism behind rate formation in the tiering system seems to be similar to our simple model.

Figure 4: Comparison between market and theoretical rates

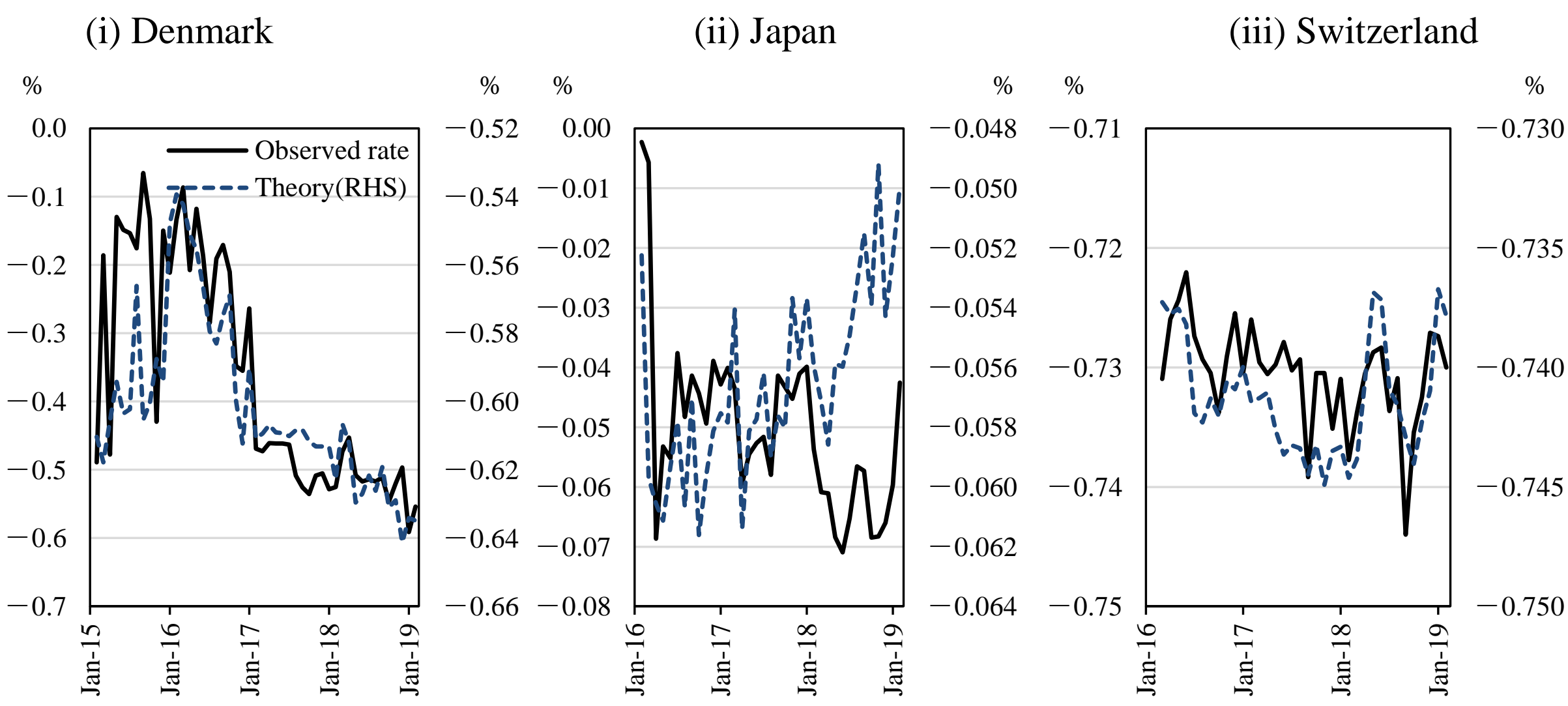

Sources: NASDAQ, Six Swiss Exchange, Danmarks Nationalbank, BoJ, SNB

Because we used the model (8) in Figure 4, the remaining gap comes from assuming $c_i \sim (0,1)$. In the case of Japan, where a rich set of information on the interbank market is available, some types of bank are said to be structurally not active in the interbank market (Bank of Japan, 2017), meaning this type of bank has the highest cost, $c_i = 1$. If we only use relevant sectors active in the interbank market to calculate $B/A$, Figures 3 and 4 would have the better fit in Figure 5.

Figure 5: Using Japanese data on relevant sectors

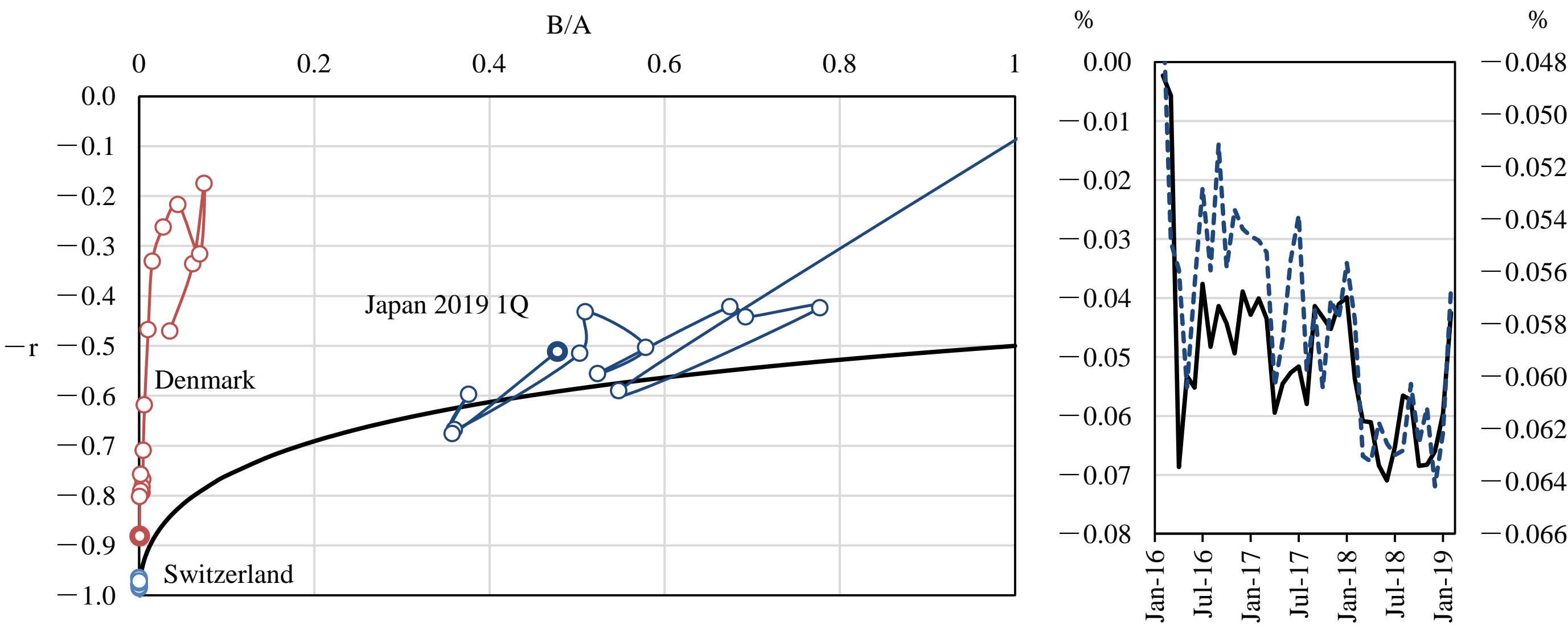

Note: City, regional, and trust banks and dealers, which are the relevant sectors in the interbank market, as specified by the Bank of Japan (2017), are considered.
Sources: NASDAQ, Six Swiss Exchange, Danmarks Nationalbank, BoJ, SNB

## 4. Application

Further, the possibility of using a tiering system for various sets of interest rates and its consequences is explored. The results are relevant under long-term large liquidity in the market. Specifically, as long as liquidity remains high, there will be low demand for interbank borrowing and no substantial increase in interbank trading volume. However, this can be overcome, in particular, for the central banks that have already utilized tiering remuneration in the negative interest rate, by keeping tiering remuneration.

As such, the remuneration for the reserve below the threshold should be higher than the remuneration exceeding the threshold. For example, remuneration schedule (12) is inversing remuneration (1), where 0% is applied to the reserve position up to $u$ and 1% is applied to the exceeding amount, which is unrealistic.

$$p_i = \begin{cases} 1 \cdot (x_{\mathrm{a}i} - u), x_{\mathrm{a}i} > u \\ 0, u \geq x_{\mathrm{a}i} \end{cases}. \quad (12)$$

This is because a bank whose position is exceeding $u$ tries to indefinitely borrow money from banks whose position is under $u$. Therefore, we generalize our model by stating that $j + 1\%$ is applied to the reserve position up to $u$ and $j\%$ to the exceeding amount. One can consider the case where the difference between the two remuneration rates is not exactly 1%, that is, by multiplying some number to the model, as we did in the comparison of our model with market data.

$$p_i = \begin{cases} j \cdot (x_{\mathrm{a}i} - u) + (j+1) \cdot u, x_{\mathrm{a}i} > u \\ (j+1) \cdot x_{\mathrm{a}i}, u \geq x_{\mathrm{a}i} \end{cases}. \quad (13)$$

Then, by stating the market rate as $j + 1 - r$ (%), the trading decision can be described as follows. Note that (14) is identical to (2). Therefore, (3)~(9) hold.

$$x_{\mathrm{a}i} = \begin{cases} x_i - l_i, x_i > u \text{ and } (j+1-r) - c_i - j > 0 \\ x_i + b_i, u > x_i \text{ and } j + 1 - (j+1-r) - c_i > 0. \\ x_i, \text{otherwise} \end{cases} \quad (14)$$

**Maximization of trading volume:** Our model helps solve the maximization of trading volume. The underlying rationale is that it is important to set an appropriate exemption for higher interbank trading. From (5) and $x_i \sim (0,1)$, the following maximization problem for $V$ can be set:

$$\text{argmax}_u V = \text{argmax}_u \frac{u^2(1-u)^2}{u^2+(1-u)^2} = \frac{1}{2}, \tag{15}$$

$$u = \frac{1}{2} \Leftrightarrow ES = 1 \Leftrightarrow r^* = 0.5. \tag{16}$$

Even if $u$ is unobservable, checking if the market rate satisfies $r = 0.5$, that is, the midpoint of two remuneration rates, is an indirect tool to discuss the maximization of interbank trading volume. Figure 6 shows the relationship between $u, 1 - r^*, V$. We use $1 - r^*$ to describe the idea of the development of the theoretical market rate; in other words, we assume $j = 0$ in the following discussion for illustrative purposes. Therefore, 0 represents the lower remuneration rate and 1 the higher remuneration rate. For example, the model in previous section represents the case of $j = -1$, and the theoretical rate in Figure 6 should be interpreted as the level which is 1 (%) lower than the written figure.

Figure 6: Simulation with upper limit of the reserve for a higher remuneration rate

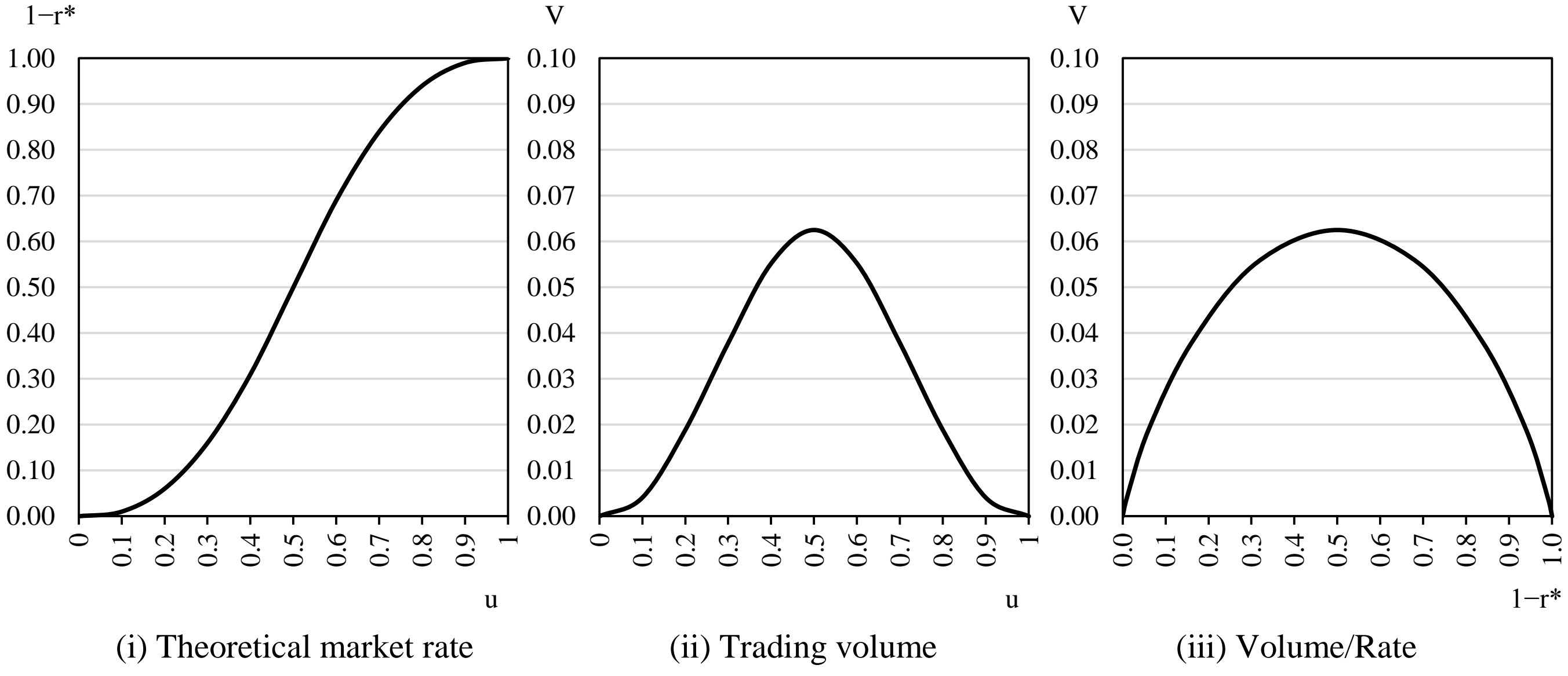

(i) Theoretical market rate (ii) Trading volume (iii) Volume/Rate

**One more tier:** After setting upper limit $u$, one might assume that it is possible to add lower limit $\ell(< u)$ to the amount of the reserve to which the higher rate $j + 1\%$ is applied. As such, the reserve outside of $[\ell, u]$ is remunerated at $j\%$.

$$p_i = \begin{cases} j \cdot (x_{\mathrm{a}i} - u + \ell) + (j+1) \cdot (u - \ell), x_{\mathrm{a}i} > u \\ j \cdot \ell + (j+1) \cdot (x_{\mathrm{a}i} - \ell), u \geq x_{\mathrm{a}i} > \ell \\ j \cdot x_{\mathrm{a}i}, \ell \geq x_{\mathrm{a}i} \end{cases}. \tag{17}$$

However, this complicates bank trade decisions. Each bank $i$ will compare the following strategies and choose the one that maximizes profit.

Strategy 1: $x_{\mathrm{a}i} = u$ ($l_i = x_i - u$ when $x_i \geq u$, $b_i = u - x_i$ when $u > x_i$)

$$\pi_{1i} = \begin{cases} (j+1-r) \cdot (x_i - u) - (j + c_i) \cdot (x_i - u), x_i > u \\ (j+1) \cdot (u - x_i) - (j+1-r+c_i) \cdot (u - x_i), u \geq x_i > \ell \\ (j+1) \cdot (u - \ell) - (j+1-r+c_i) \cdot (u - x_i), \ell \geq x_i \end{cases}. \tag{18}$$

Strategy 2: $x_{\mathrm{a}i} = 0$ ($l_i = x_i$)

$$\pi_{2i} = \begin{cases} (j+1-r) \cdot x_i - (j + c_i) \cdot (x_i - u + \ell) - (j+1+c_i) \cdot (u - \ell), x_i > u \\ (j+1-r) \cdot x_i - (j + c_i) \cdot \ell - (j+1+c_i) \cdot (x_i - \ell), u \geq x_i > \ell \\ (j+1-r) \cdot x_i - (j + c_i) \cdot x_i, \ell \geq x_i \end{cases}. \tag{19}$$

Strategy 3: $x_{\mathrm{a}i} = x_i$ (doing nothing)

$$\pi_{3i} = 0. \tag{20}$$

The strategy choice above by banks is described as

$$s_i = \{1, 2, 3\}. \tag{21}$$

As before, $\mathrm{r}^*$ should be decided to satisfy (3):

Using this lower limit model, cases $\ell = 0.4 \times u$ and $\ell = 0$ were compared in the same setting in Figure 5. In Figure 6, $1 - r^*, V$, and $P$ [described in (22)] were plotted over $u$, assuming $j = 0$ again. The underlying reason for discussing $P$ here is that setting the lower limit will reduce the payment from the central bank to commercial banks:

$$P = \int_i \; p_i \, \mathrm{d}i. \tag{22}$$

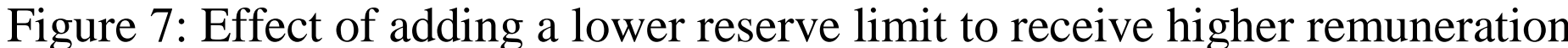
Figure 7: Effect of adding a lower reserve limit to receive higher remuneration

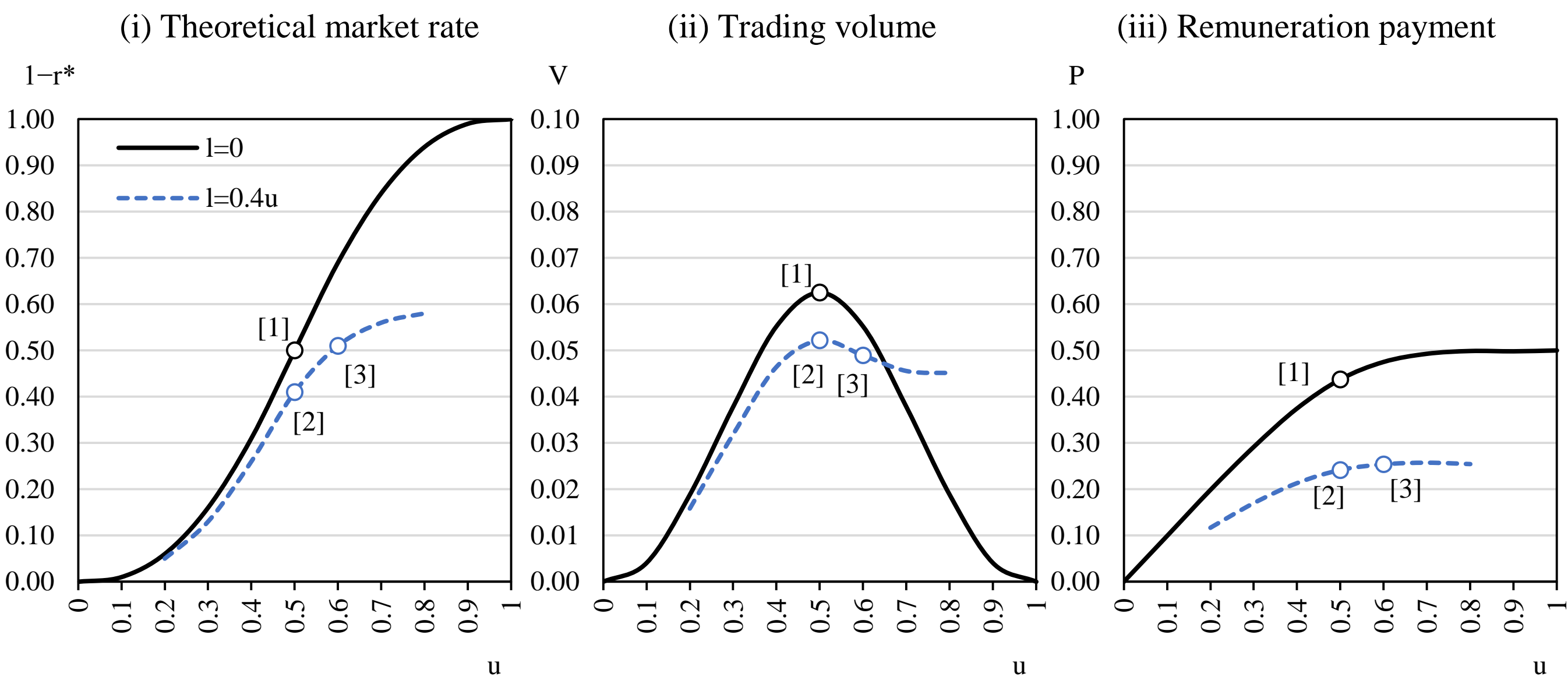

Figure 7 shows that adding $\ell$ decreases $1 - r^*$, $V$ (if $u$ is around 0.5), and $P$. Particularly, we start from [1] without lower limit, where the central bank sets $(u, \ell) = (0.5, 0)$ with $1 - r^* = 0.5$ and maximized $V$. Next, we move to [2] by adding lower limit $\ell = 0.2$. Specifically, $(u, \ell) = (0.5, 0.2)$ and $1 - r^* = 0.41$, and $P$ declines by 45% while $V$ only decreases by 17%. Additionally, in [3], the central banks that want to keep $1 - r^* = 0.5$ may choose $(u, \ell) = (0.6, 0.24)$. In that case, the central banks could reduce their interest payment by 42% if they risk decreasing the market trading volume by 22% without changing the market rate. This is based on comparing [1] with [3].

This study does not aim to advise central bankers how to save interest payments to the deposits of commercial banks. In fact, adopting a tiering remuneration system will not save money for the central banks. Figure 8 illustrates the impact of adopting a tiering system by starting from case [0], which is not a tiering system, particularly $(u, \ell) = (1, 0)$ and then $1 - r^* = 1.0$. When a central bank adds an upper limit, the point will move to [1] and $(u, \ell) = (0.5, 0)$. As desired, $V$ will significantly increase but $1 - r^*$ will decline to $1 - r^* = 0.5$. Under this condition, central banks should double the remuneration rate if they want to maintain the market rate at 1.0, and then the interest payment calculated in the right-hand side of Figure 7 will double. Therefore, this necessary adjustment for a higher remuneration is incorporated in Figure 8 by $P/(1 - r^*)$. Adding a lower limit leads to [3] and $(u, \ell) = (0.6, 0.24)$, where adjusted $P$ is similar to [0]. Therefore, there is no effect of saving on interest payment. As always, there is no free lunch for central banks (i.e., interest payment on reserves).

Figure 8: Effect of adopting a tiering remuneration system

(i) Theoretical market rate

(ii) Trading volume

(iii) Remuneration payment (after remuneration adjusted)

**Trade-offs:** To visualize the trade-off between trading volume and remuneration payment, we would draw their relationship for possible pairs of $(u, \ell)$. Specifically, $u = 0.01, 0.02, \cdots, 1.00$ and $\ell = 0, 0.01, \cdots, 0.99$ satisfy (23) and (24)

$$u > \ell, \tag{23}$$

$$\int_{s_i=1} l_i > \int_{s_i=2} l_i. \tag{24}$$

Condition (24) is set to have a realistic choice of $(u, \ell)$. Otherwise, the primary lenders in the market would be banks who have less liquidity ($s_i = 2$) and the primary borrowers would be the banks with relatively higher liquidity. To avoid such polarization, we assume condition (24) holds.

Figure 9 shows the set of trading volume and remuneration payment after adjusting the level of remuneration, assuming $j = 0$.

Figure 9: Trade-off of volume and cost saving for a central bank

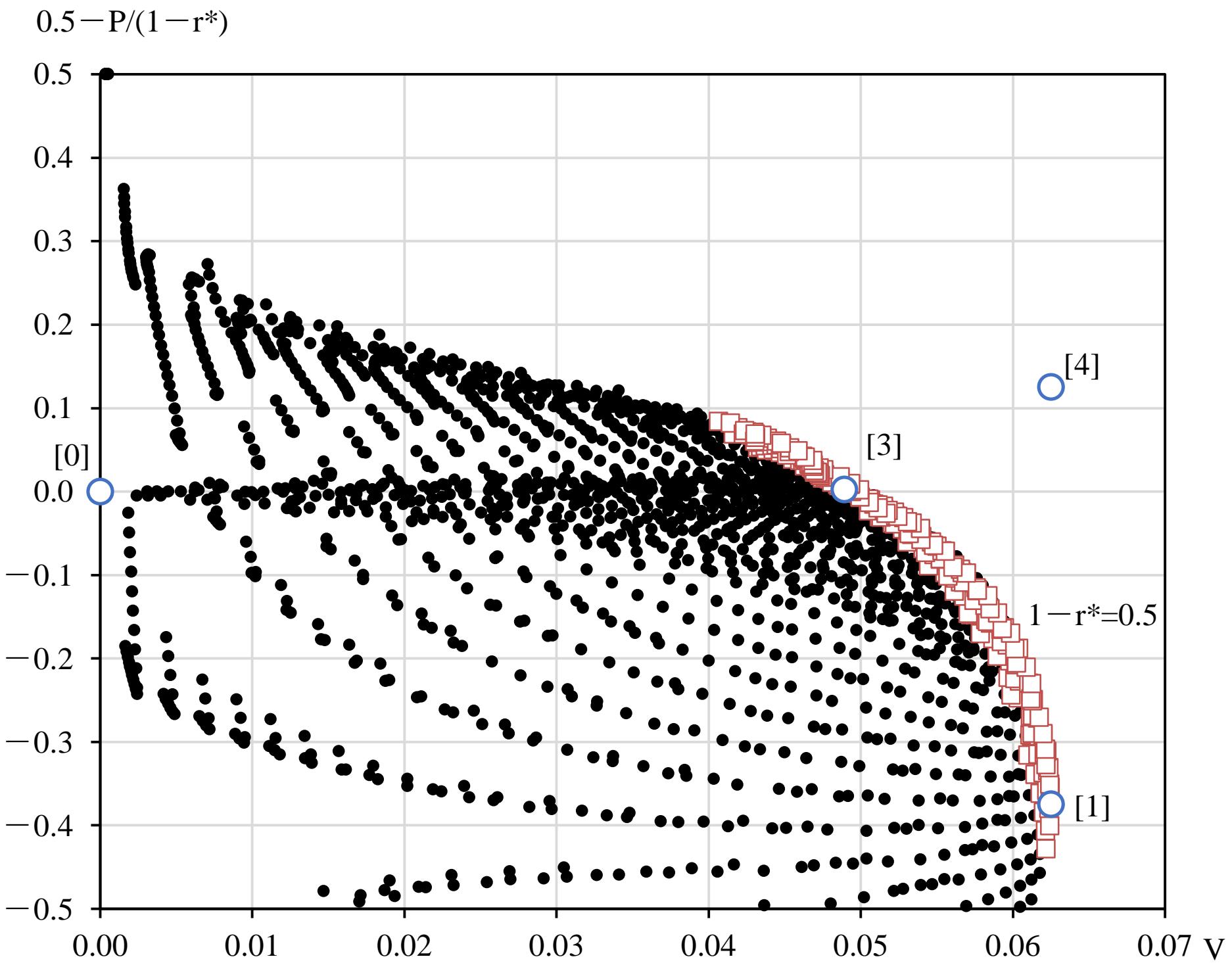

The vertical axis in Figure 9 shows how much a central bank can save compared to situation [0]: $(u, \ell) = (1, 0)$ without a tiering system. The horizontal axis shows the trading volume. Therefore, the upper right side is preferable for the central bank. Further, [1] and [3] are the same as before, so both $(u, \ell) = (0.5, 0)$ and $(u, \ell) = (0.6, 0.24)$ are optimal settings, that they are on the upper right corner of the possible domain. In fact, the pair satisfies $1 - r^* = 0.5$, being shown as squares in Figure 9, are all around on the margin in the upper right corner. Therefore, a first-hand implication from our model is that central banks can try to set tiering remuneration limit to realize the market rate around the midpoint of remuneration rates.

However, once we drop assumption (25), point [4]: $(u, \ell) = (1.0, 0.5)$ will be an option, where the trading volume is maximized and the central bank can save remuneration payment at the same time. If central banks do not regard bank polarization as a problem, they can choose this solution. We will not discuss the legitimacy of this approach, but it is worth mentioning that it implies the maximization of trading volume is accompanied by a side effect of either the increase of remuneration cost [1] or polarization of bank reserves [4].

So far, we have focused on cost for central banks, but the reduction in the remuneration payment means a decline of income for commercial banks. This viewpoint is essential especially when a negative rate is imposed on reserves, because the remuneration payment is the cost for commercial banks. If $j = -1$, then the shape of the curve in Figures 6 and 7 holds, while the theoretical rate should be read as the level which is 1 (%) lower than the written figure, and remuneration cost should be interpreted as inversed. Specifically, the higher the remuneration payment on the right-hand side chart in Figure 7 becomes, the higher the exemption of the negative remuneration applied to commercial banks is. Consequently, the possible domain of trading volume and negative remuneration cost for commercial banks is shown in Figure 10, as the upper right side is preferable for commercial banks.

Figure 10: Trading volume and cost for commercial banks under negative remuneration

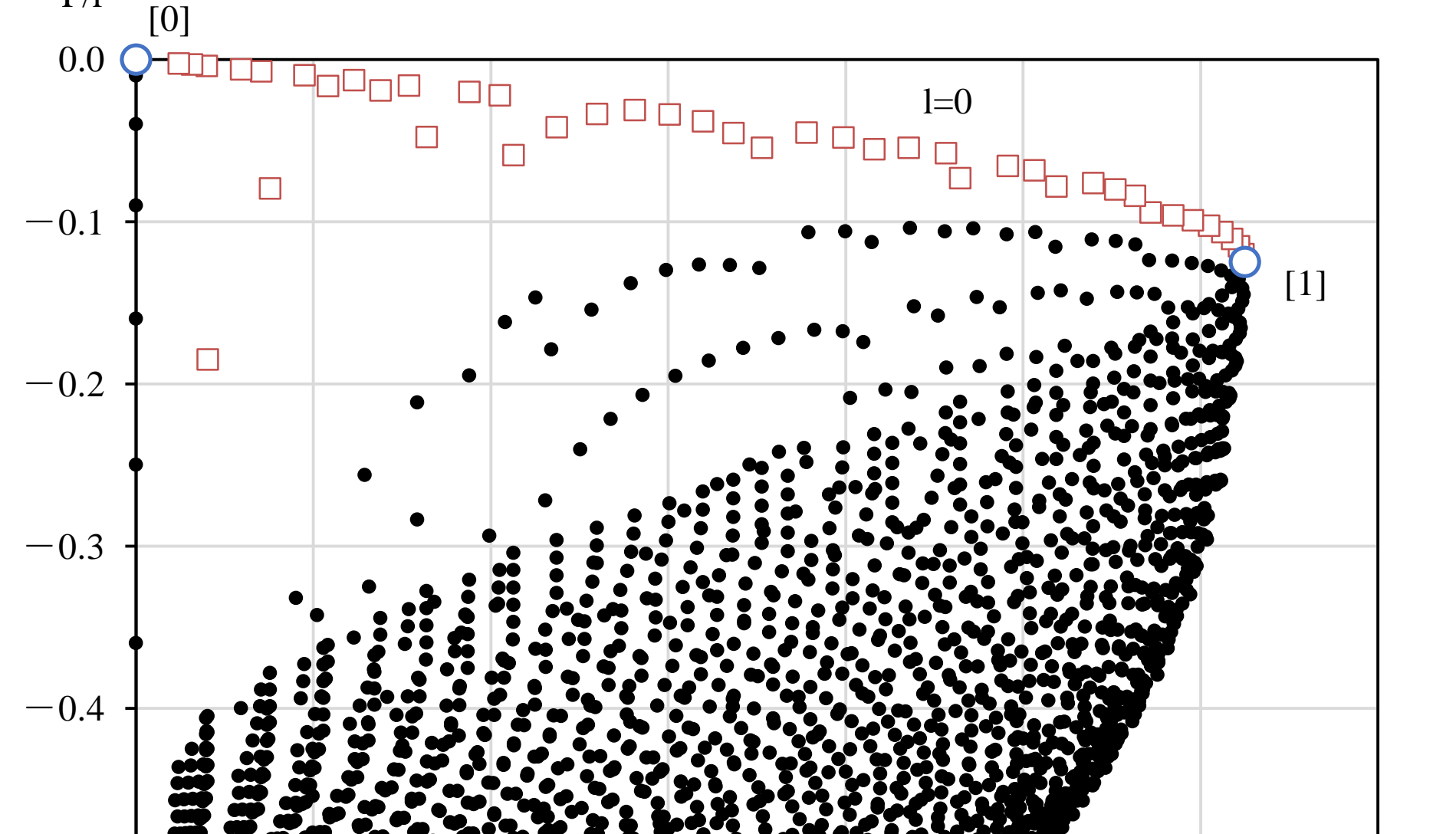

The interpretation is similar to Figure 9. The vertical axis in Figure 10 shows the negative rate charged to commercial banks. The horizontal axis shows the trading volume. Then, [1]: $(u, \ell) = (0.5, 0)$ is considered as a good choice, but setting lower limit will worsen condition [3]: $(u, \ell) = (0.6, 0.24)$. Instead, the small squares, which are the set of $(u, \ell)$ satisfying $u > 0.5, \ell = 0$, are on the margin in the upper right corner. That is, in a negative rate environment, central banks do not have to adopt a lower limit for a reserve to receive higher remuneration by keeping in mind the cost for commercial banks.

## 5. Conclusions

This study proposes a model of interbank market with tiering remuneration. This model was simplified by using only limited inputs to predict the interbank market rate: (i) aggregated reserve exceeding higher remuneration limit and (ii) unutilized amount below higher remuneration limit, or (iii) exemption share over the excess liquidity in the system. The prediction of real market data was not perfect, but acceptable. Therefore, further research should consider a better model fit. The model in this study has a good balance of simplicity and performance.

By expanding this model, the applications of tiering remuneration were also discussed. Two key conclusions may be drawn from the model. First, setting an upper limit for receiving remuneration for reserves will enhance interbank trade, and this trading volume will be maximized when the market rate falls at the midpoint of the two remuneration rates. However, a central bank's payment for interest on reserves will also increase. That is, volume increases when sacrificing the central bank's interest payments.

Second, adding a lower limit of receiving remuneration for reserves will help the central bank offset the increase in the remuneration payment discussed earlier. Moreover, the reduction in trading volume is relatively limited. Therefore, using both the upper and lower limits for a reserve to receive higher remuneration will increase market volume, whereas the market rate and the amount of remuneration will not change. Therefore, central bankers, especially those who have already adopted a tiering system, may consider maintaining the tiering interest rate to retain an active interbank market.